\documentclass[twocolumn,showpacs,preprintnumbers,amsmath,amssymb,psfig]{revtex4}

\usepackage{graphicx}
\usepackage{dcolumn}
\usepackage{bm}

\newcommand{\beq}{\begin{eqnarray}}
\newcommand{\eeq}{\end{eqnarray}}
\newcommand{\beqq}{\begin{eqnarray*}}
\newcommand{\eeqq}{\end{eqnarray*}}

\begin{document}
\title
{Potential Barrier Classification by Short-Time Measurement}
\author{Er'el Granot}
 \email{erel@yosh.ac.il}
 \affiliation{Department of Electrical and Electronics Engineering, College of Judea and Samaria, Ariel, Israel\\
}
\author{Avi Marchewka}%
 \email{marchew@post.tau.ac.il}
\affiliation{Kibbutzim College of Education\\
Ramat-Aviv, 104 Namir Road 69978 Tel-Aviv, Israel\\
\\
}
\date{\today}
\begin{abstract}
We investigate the short-time dynamics of a delta-function
potential barrier on an initially confined wave-packet. There are
mainly two conclusions: A) At short times the probability density
of the first particles that passed through the barrier is
unaffected by it. B) When the barrier is absorptive (i.e., its
potential is imaginary) it affects the transmitted wave function
at shorter times than a real potential barrier. Therefore, it is
possible to distinguish between an imaginary and a real potential
barrier by measuring its effect at short times only on the
transmitting wavefunction.
\end{abstract}
\pacs{03.65.-w, 03.65.Nk, 03.65.Xp.}
 \maketitle
\section{Introduction}
One of the methods in imaging a turbid or a diffusive medium with
optical radiation is the time-gating technique
\cite{TG1,TG2,TG3,TG4,TG5}. In this technique a temporally narrow
pulse is injected into the medium. Owing to the diffusivity of the
medium, when the pulse exits the medium it becomes considerably
wider. However, if the first arriving photons are separated from
the rest of the pulse then it is possible to use these, so called,
ballistic (or quasi ballistic) photons to reconstruct the
ballistic image of the medium.

Therefore, by employing a short time-gating technique the
multi-scattering effect can be eliminated. Indeed, such methods
were employed in recognizing hidden objects and informative
signals in diffusive media \cite{TGE1}. Naively, one might expect
that this technique can be implemented for electron imaging to
"see" absorptive objects inside scattering medium. That is, one
can send a short pulse of electrons to one end of the medium,
while at the other end only the first arriving electrons will be
measured. By doing so all the noise caused by the multi-scattering
should be eliminated. On the other hand, the presence of
absorptive regions (like imaginary potentials \cite{Muga_PR}) will
be felt in the amount of the early arriving electrons. However,
electrons are governed by the Schr\"{o}dinger equation, and unlike
the Maxwell wave-equation, has a parabolic dispersion relation. As
a consequence, any localized wave-packet suffers from strong
dispersion, since each spectral component propagates at different
velocity. The fastest particles are the most energetic ones, which
pass through the medium unaffected, since the barrier's potential
energy is negligible compared to their kinetic energy. In
particular, when the medium is a certain barrier (or well), we
conclude that when only the first-arriving particles are measured
there should be no trace of the barrier's presence. It does not
matter what shape or height the barrier has - the first particles
that pass though the barrier should be indifferent to it. Thus,
one may argue that the time-gating technique cannot be implemented
to electrons imaging, at least not in its naive form. However, we
show that the short-time measurement reveals information about the
{\emph nature} of the barrier-- whether it is imaginary or real.

There is a peculiar distinction between an absorbing medium (e.g.,
an imaginary potential) and a non-absorbing one (e.g., a real
potential). While they both have no effect on the wavepacket (both
transmitted and reflected) at $t \rightarrow 0$, the imaginary
barrier influences the wavepacket sooner. In other words, in the
temporal Taylor expansion of the probability density the imaginary
potential appears at smaller order than a real potential. It is
then clear that we can classify the barrier as an absorptive one
simply by measuring the wave-packet at short times. Note that in
general it is required to measure both reflection and transmission
coefficients to figure out if the barrier is absorptive or not.

Recently \cite{Fort_PRL}, it has been demonstrated even
experimentally that it is feasible to investigate the 1D
scattering of a Bose-Einstein condensate by a narrow defect.
Therefore, it seems that there is a good chance of witnessing
these effects in the laboratory in the near future.

In this paper we demonstrate this effect rigorously (both
analytically and by a numerical simulation) for the delta function
potential. That is, we show that it is possible to identify an
absorptive potential by measuring the short time dynamics of only
the transmitted wavefunction.

The initial state we consider is a wave-packet, which is confined
to one side of the barrier. It is then demonstrated that the
wavefunction at the other side is independent of the barrier for
short times, while the temporal dependence depends on the exact
nature of the barrier (absorptive or not).

\section{System description and dynamics}

Evidently, in order to confine the initial wavepacket to one side
of the barrier, there has to be a certain singularity in the
wavepacket. In this paper we focus on a step function to simplify
the problem; however, it has been demonstrated elsewhere that most
of the conclusions are valid even in the continuous case provided
the measurements are taken at specific ranges (see ref.\cite{us}).
Is is also demonstrated at the end of the paper that the main
conclusions are valid even when the initial wavepacket is a
Gaussian. For simplicity we take a 1D delta function as the
potential barrier.

The system illustration is depicted in fig.1. Initially, the wave
packet has the form \cite{Muga_PR,us,m}
\begin{equation}\label{}
  \psi(x,t=0)=\theta \left(-x\right)\exp\left(ik_0x\right)
\end{equation}

and a distance $L$ from its front we place a delta-function
barrier $V(x)=\lambda \delta (x-L)$ (see fig.1). That is, the
Schr\"{o}dinger equation reads
\begin{equation}\label{}
   -\frac{\partial ^2}{\partial x^2}\psi+\lambda \delta(x-L)\psi=
   i\frac{\partial\psi}{\partial t}
\end{equation}

\begin{figure}
\includegraphics[width=8cm,bbllx=70bp,bblly=635bp,bburx=390bp,bbury=780bp]{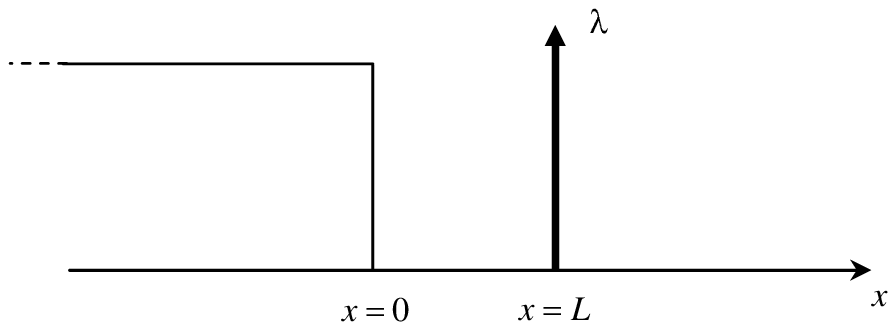}
\caption{\emph{System schematic: a semi-infinite wavepacket
hitting a delta function potential.}}\label{fig1}
\end{figure}

therefore for $t>0$
\begin{equation}\label{}
  \psi(x,t)=\frac{1}{2\pi}\int dk\varphi(k)\chi(k,x)\exp(-ik^2t)
\end{equation}
where
\begin{equation}\label{}
\varphi(k)=\frac{i}{k-k_0+i0}
\end{equation}
is the Fourier transform of $\psi(x,t=0)$ and
\begin{equation}\label{}
\chi(k,x)\equiv\exp(ikx)+\frac{i\lambda /2}{k-i\lambda
/2}\exp\left(ik|x-L|+ikL\right)
\end{equation}
The general solution is (for $x>0$)
\begin{widetext}
\begin{eqnarray}
\psi(x,t)=\frac{e^{ix^2/4t}}{2}w\left[\sqrt{it}\left(\frac{x}{2t}-k_0\right)\right]+
\frac{e^{iy^2/4t}}{2} \frac{i\lambda /2}{k_0-i\lambda/2}\
 \left\{w\left[\sqrt{it}\left(\frac{y}{2t}-k_0\right)\right]
   -w\left[\sqrt{it}\left(\frac{y}{2t}-i\frac{\lambda}{2}\right)\right]\right\}
 \end{eqnarray}
where $w(z)\equiv \exp(-z^2)\mbox{erfc}(-iz)$ \cite{AS} and
$y\equiv L+|x-L|$. In short times one can expand this expression
in powers of $t$. Up to $t^{5/2}$
\begin{eqnarray}
 \psi(x,t)&\simeq &\sqrt{\frac{it}{\pi}}\frac{e^{ix^2/4t}}{x}
  \left[1+\frac{2\left(k_0x-i\right)}{x^2}t+\frac{4\left(k^2x^2-3ik_0x-3\right)}{x^4}t^2\right]
  +\frac{(it)^{3/2}}{\sqrt{\pi}}\frac{e^{iy^2/4t}}{y^2}\lambda
\left[1+\frac{i(\lambda y-6)+2k_0y}{y^2}t\right]
 \end{eqnarray}
  for $x>L$
\begin{eqnarray}
 \psi(x,t)&\simeq& \sqrt{\frac{it}{\pi}}\frac{e^{ix^2/4t}}{x}\nonumber
  \left[1+\frac{2(k_0x-i)}{x^2}t+
  \frac{4(k^2x^2-3ik_0x-3)}{x^4}t^2+
  \frac{i\lambda}{x}t-\frac{\lambda(\lambda x-6-2ik_0x)}{x^3}t^2\right]
 \end{eqnarray}
 \end{widetext}
  and to the third order of $t$:
\begin{eqnarray}
\label{third}
&&|\psi(x>L,t)|^2\simeq \frac{t}{\pi
 x^2}\times \nonumber
 \\
 &&\left\{1+4\frac{k_0t}{x}+4\frac{3(k_0x)^2-5}{x^4}t^2+
 \frac{\lambda}{x^2}t^2(\frac{8}{x}-\lambda)\right\}
\end{eqnarray}

 we can see that the barrier's presence is felt only at the third order of
 $t$.\\
 Even when $k_0\ll\lambda$ and $x \rightarrow \infty$ the barrier's presence
 has a significant influence when the measurement is taken in the
 range $4k_0x/{\lambda^2}\ll t\ll x/\lambda$.\\
 On the other hand, for $0<x<L$, owing to the reflection from the barrier,
 the probability density's dependence on the barrier appears even in
 the $t^{3/2}$ order.\\
\begin{eqnarray}\label{}
\psi(x,t)&\simeq& \sqrt{\frac{i}{\pi}}
\frac{e^{ix^2/4t}}{x}\left[t^{1/2}+\frac{2t^{3/2}}{x}(k_0+\frac{i}{x})\right]\nonumber
\\
&&+i\lambda\sqrt{\frac{i}{\pi}}\frac{e^{i(2L-x)^2/4t}t^{3/2}}{(2L-x)^2}
\end{eqnarray}
and
\begin{eqnarray}\label{}
&&|\psi(x,t)|^2\simeq \frac{t}{\pi}\times \\
 &&\left\{
\frac{1}{x^2}\left[1+\frac{4tk_0}{x}\right]-\frac{2\lambda
t}{x(2L-x)^2}\sin \left[\frac{L(L-x)}{t}\right]\right\}\nonumber
\end{eqnarray}
That is, the dependence appears at the probability density at
the coefficient of $t^2$.\\
Obviously, this approximation applies only when the argument
$L(L-x)/t$ is not too small. \\

\begin{figure}
\includegraphics[width=8cm,bbllx=30bp,bblly=170bp,bburx=580bp,bbury=620bp]{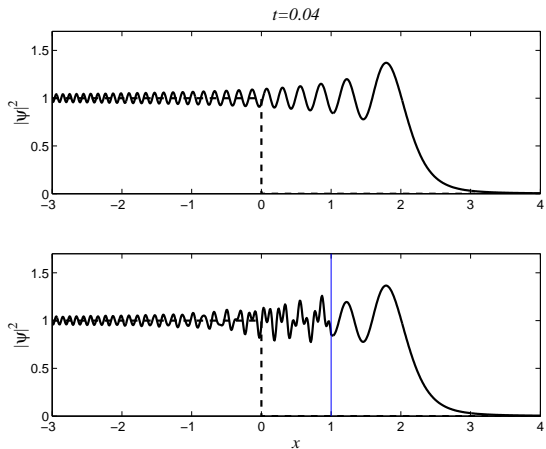}
\caption{\emph{A comparison between the solution with (lower
panel) and without (upper panel) the barrier, which is represented
by the horizontal line at $x=1$. The initial state is represented
by the dashed line. The parameters in this case are $L=1$,
$k_0=30$, $t=0.04$ and $\lambda=3$.}}\label{fig2}
\end{figure}

In Fig.2 we plot a comparison between the propagation of the
wavepacket in case the barrier is absent (upper panel) and when it
is present (lower panel). In Fig.3 the difference between the two
(with and without the barrier, i.e.,
$\Delta|\psi|^2=|\psi|_{with}^2-|\psi|_{without}^2$) is plotted.
Despite the fact that the packet passes {\em through} the
potential, its effect beyond the barrier $x>L$ is miniscule and
for $|x|^2\gg 1$ the two solutions are essentially identical.
Moreover, the difference between the $x>L$ and $x<L$ regimes is
clear from the figure. In the latter regime the influence of the
potential is felt for longer distances, but still when
$\frac{|x|^2}{t}\rightarrow \infty$ its influence decays to
zero.\\
 \\

\begin{figure}
\includegraphics[width=8cm,bbllx=30bp,bblly=170bp,bburx=580bp,bbury=620bp]{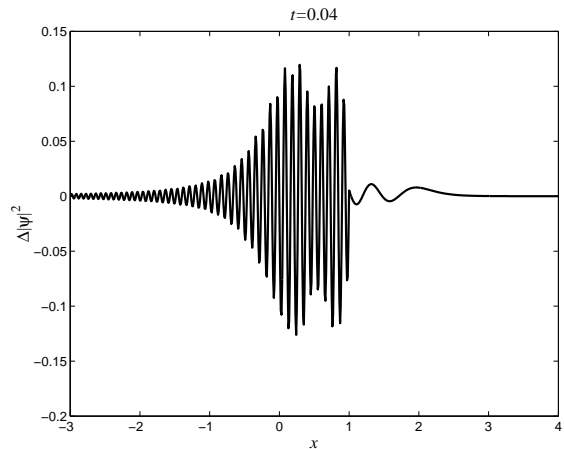}
\caption{\emph{: The difference between the solution when the
barrier is present and when it is absent. Clearly, as $|x|^2/t\gg
1 $ the two solutions are identical.}}\label{fig3}
\end{figure}

When the potential is absorptive the Schr\"{o}dinger equation
should be rewritten
\begin{equation}\label{}
   -\frac{\partial ^2}{\partial x^2}\psi+i\lambda \delta(x-L)\psi=
   i\frac{\partial\psi}{\partial t}
\end{equation}
\begin{equation}\label{}
\psi(x,t)=\sqrt{\frac{it}{\pi}}
\frac{e^{ix^2/4t}}{x}\left[1+\frac{2(k_0x-i)-\lambda
x}{x^2}t+O\left(t^2\right)\right]
\end{equation}
and the probability density satisfies
\begin{equation}\label{}
|\psi(x>L,t)|^2={\frac{t}{\pi x^2}}
\left\{1+2\frac{2k_0-\lambda}{x}t+O(t^2)\right\}.
\end{equation}
In this case the potential presence appears in the probability
density at the second order of $t$ (and not the third as in eq.\ref{third}).\\
In the case where the wavefunction initially vanishes at $x=0$
 (\cite{us,m,g}), e.g., when
\begin{equation}\label{}
\psi(x,t=0)=\theta (-x)\sin (k_0x)
\end{equation}
The general solution is
\begin{widetext}
\begin{eqnarray}
\psi(x,t)&=&\frac{e^{ix^2/4t}}{4i}w\left[\sqrt{it}(\frac{x}{2t}-k_0)\right]
+\frac{e^{iy^2/4t}}{4} \frac{\lambda /2}{k_0-i\lambda /2}
\left\{w\left[\sqrt{it}(\frac{y}{2t}-k_0)\right]
-w\left[\sqrt{it}(\frac{y}{2t}-i\frac{\lambda}{2})\right]\right\}\\
&&
-\left\{\frac{e^{ix^2/4t}}{4i}w\left[\sqrt{it}(\frac{x}{2t}+k_0)\right]-
\frac{e^{iy^2/4t}}{4} \frac{\lambda /2}{k_0+i\lambda/2}
\left\{w\left[\sqrt{it}(\frac{y}{2t}+k_0)\right]
-w\left[\sqrt{it}(\frac{y}{2t}-i\frac{\lambda}{2})\right]\right\}\right\}\nonumber
\end{eqnarray}
\end{widetext}
In short times one can expand this expressions in powers of $t$.\\
Up to $t^{5/2}$
\begin{eqnarray}\label{}
\psi(x,t)&\simeq& -\sqrt{\frac{i}{\pi}}\frac{2k_0
t^{3/2}e^{ix^2/4t}}{x^2}\left[i+\frac{6}{x^2}t\right]\\
&+&\sqrt{\frac{i}{\pi}}\frac{2k_0 t^{5/2}e^{iy^2/4t}}{y^3}\lambda
\nonumber
\end{eqnarray}
for $x>L$
\begin{equation}\label{}
\psi(x,t)\simeq \frac{t^{3/2}}{\sqrt{i\pi}}
\frac{e^{ix^2/4t}}{x^2}2k_0\left[1+\frac{it}{x}\left(\lambda-\frac{6}{x}\right)\right]
\end{equation}
and
\begin{equation}\label{}
|\psi|^2\cong \frac{4k_0^2t^3}{\pi x^4}+O\left(t^4\right)
\end{equation}
we can see that the barrier is not felt even at the third order of
$t$.\\
Although the phase of $\psi$ , "feels" the barrier even in the
first order of $t$, the leading term is proportional to $t^{-1}$
\begin{equation}\label{}
\angle \psi\cong -\frac{\pi}{4}+\frac{x^2}{4t}+\left(x\lambda
-6\right)\frac{t}{x^2}.
\end{equation}

\section{The continuous case: Gaussian dynamics}

The fact that we used singular initial wavefunction may raise
skepticism about the physical validity of the conclusions.
However, the main conclusions of the semi-infinite plane wave can
be deduced even when the initial wavepacket is a continuous
wavefunction, such as a Gaussian. However, owing to the finite
spectral width of the Gaussian, high energy particles are very
rare in the packet, and therefore the reasoning that was used in
the semi-infinite plane wave can be used only in the intermediate
temporal region and fails at $t\rightarrow 0$. Therefore, we
should expect to find the same conclusions in a certain
intermediate period (as in ref. \cite{us}).
\\

If the initial wavepacket is a Gaussian:

\begin{equation}\label{}
\psi
(x,t=0)=\sqrt{\frac{2}{\pi}}\frac{1}{\sigma}\exp\left\{-(\frac{x}{\sigma})^2+ik_0x\right\}
\end{equation}
then
\begin{eqnarray}\label{}
&&\psi (x>L,t\geq 0)=\\
&&\frac{1}{\sqrt{2}\pi}\int dk
\frac{\exp\left\{-\left(k-k_0\right)^2\sigma
^2/4+ikx-ik^2t\right\}}{1-i\lambda/2k}\nonumber
\end{eqnarray}
the general solution is
\begin{widetext}
\begin{eqnarray}
 \psi (x>L,t\geq
0)=\left\{\frac{1}{\sqrt{2\pi}s}-\frac{\lambda}{\sqrt{8}}w\left[i\frac{\lambda
s }{2}-\frac{\left[\sigma^2 k_0/2+ix\right]}{2s}\right]\right\}
\exp\left\{\frac{1}{4}\frac{[\sigma^2
k_0/2+ix]^2}{s^2}-\frac{\sigma^2k_0^2}{4}\right\}
\end{eqnarray}
where $s\equiv \sqrt{\sigma^2/4+it}$. For short times
\begin{eqnarray}
 \psi (x>L,t\geq 0)\simeq
\frac{1}{\sqrt{2\pi}s}\frac{1}{1-i\frac{\lambda s^2}{\sigma^2
k_0/2+ix}} \exp\left\{\frac{1}{4}\frac{\left[\sigma^2
k_0/2+ix\right]^2}{s^2}-\frac{\sigma^2k_0^2}{4}\right\}
\end{eqnarray}
\end{widetext}
where for large distances $x+L\gg\sigma^2k_0$ can be approximated
in the two extreme cases: $t\ll \sigma ^2/4$ and $t\gg \sigma
^2/4$. In the former case
\begin{eqnarray}
\psi (x>L,t\geq
0)=\frac{\sqrt{2}}{\sqrt{\pi}\sigma}\frac{1}{1-\frac{\lambda
\sigma^2}{4x}} \exp\left\{-\frac{x^2}{\sigma^2}\right\}
\end{eqnarray}
the barrier influence is independent of time, and in fact, very
far from the barrier  $x\gg \lambda\sigma ^2$ the barrier's
influence is negligible.  However, we see here that in the limit
$t\rightarrow 0$, due to the finite spectral width $\sim\sigma$ ,
the presence of the barrier is always felt. (Note that the
singular case $4x=\lambda \sigma^2$ is not consistent with the
above approximation). This is to be expected, since in a Gaussian
distribution the number of particles with extremely large enegies
is exponentially small.

When  $t\gg \sigma^2/4$
\begin{eqnarray}
\psi (x>L,t\geq
0)=\frac{1}{\sqrt{2{\pi}it}}\frac{1}{1-i\frac{\lambda t}{x}}
\exp\left\{\frac{i}{4}\frac{x^2}{t}-\frac{\sigma^2k_0^2}{4}\right\}\nonumber
\end{eqnarray}
we recognize a penetration velocity. When $x/t\ll\lambda$ the
barrier has a large impact, however, if the particles' velocity is
very large $x/t\gg\lambda$   the barrier's influence is
negligible.

And similarly, in the temporal period $\sigma^2/4 \ll t \ll
x/\lambda$ the difference between real and imaginary barrier is
apparent. For a real barrier

\begin{eqnarray}
|\psi (x>L,t>0)|^2\cong \frac{1-(\lambda t/x)^2}{2\pi t}
\exp\left\{-\frac{\sigma^2k_0^2}{4}\right\}
\end{eqnarray}

while for an imaginary one

\begin{eqnarray}
|\psi (x>L,t>0)|^2\cong \frac{1-2\lambda t/x}{2\pi t}
\exp\left\{-\frac{\sigma^2k_0^2}{4}\right\}
\end{eqnarray}

Again, the the influence of the absorptive potential appears in a
lower order term.

\section{Schematic experimental realization}

\begin{figure}
\includegraphics[width=8cm,bbllx=100bp,bblly=550bp,bburx=430bp,bbury=770bp]{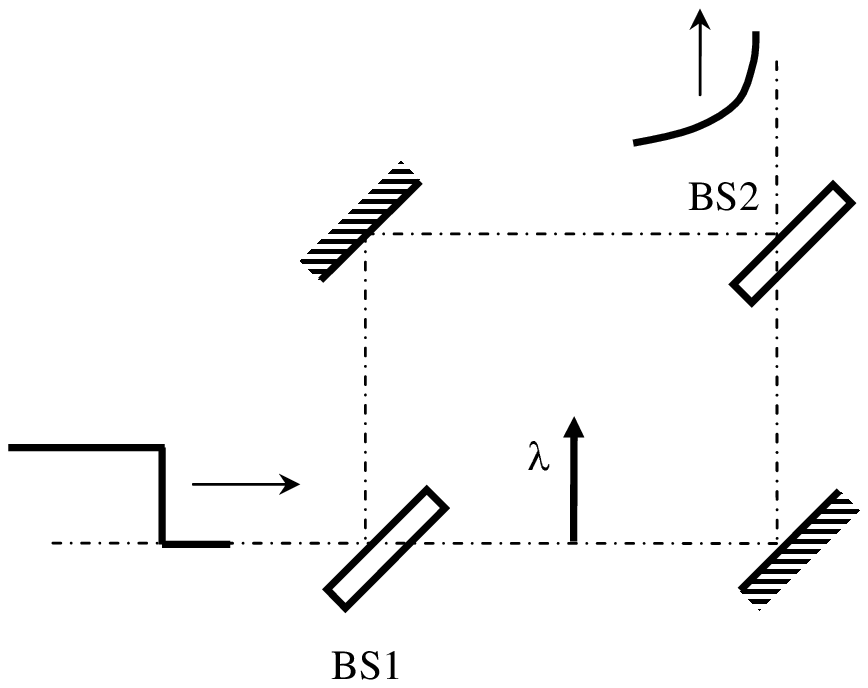}
\caption{\emph{schematic illustration of a Mach-Zehnder
interferometer for barrier classification.}}\label{fig4}
\end{figure}

One of the methods of emphasizing the impact of the potential is
by placing the potential barrier (delta function in our case) in
one arm of a Mach-Zehnder interferometer (see fig.4).

Let us denote by $c_1$  and $ic_2$  the transmission and
reflection coefficients of each of the two interferometers' beam
splitters (BS1 and BS2 in figure.4). With this notation we assume
(without loss of generality) that both $c_1$ and $c_2$ are real
and conservation of energy implies $c_1^2+c_2^2=1$.
\begin{widetext}
\begin{eqnarray}
 \psi(x,t)&=&\sqrt{\frac{it}{\pi}}\frac{e^{ix^2/4t}}{x}\left\{
  \left[1+\frac{2(k_0x-i)}{x^2}t+\frac{4(k_0^2x^2-3ik_0x-3)}{x^4}t^2\right]
  \left(c_1^2-c_2^2\right)
  +\left[\frac{i\lambda}{x}t-\frac{\lambda(\lambda
  x-6-2ik_0x)}{x^3}t^2\right]c_1^2\right\}\nonumber
 \end{eqnarray}
 \end{widetext}

Let us further assume that both BS's are almost 50:50, i.e.,
$c_1^2=0.5(1-\varepsilon),c_2^2=0.5(1+\varepsilon)$ and
$\varepsilon\ll 1$. In this case, for short time
\begin{equation}\label{}
 \psi(x,t)\simeq \sqrt{\frac{it}{\pi}}\frac{e^{ix^2/4t}}{x}
 \left(\varepsilon+\frac{i\lambda}{2x}t\right)
 \end{equation}

 and the probability density can be approximated
\begin{equation}\label{}
  | \psi(x,t)|^2\simeq \frac{t}{\pi
  x^2}\left[\left(\varepsilon-\frac{\Im\lambda}{2x}t\right)^2+(\frac{\Re
  \lambda}{2x}t)^2\right]
\end{equation}
Thus, when the potential is real the potential-dependent term has
a cubic dependence on time
\begin{equation}\label{short_real}
  | \psi(x,t)|^2\simeq \frac{t}{\pi
  x^2}\left[\varepsilon^2+\left(\frac{
  \lambda}{2x}t\right)^2\right]
\end{equation}
while if the potential is imaginary the temporal dependence of the
potential-dependent term is parabolic
\begin{equation}\label{short_imag}
  | \psi(x,t)|^2\simeq \frac{t\varepsilon}{\pi
  x^2}[\varepsilon-\frac{
  \Im \lambda}{x}t]
\end{equation}

To emphasize the difference we define $\Delta|\psi|^2 \equiv
|\psi|^2-|\psi|_{free}^2$ as the difference between the
probability density at the interferometer exit when the barrier is
present ($|\psi|^2$) and when it is absent ($|\psi|_{free}^2$).

In fig.5 we plot the temporal evolution of $\Delta|\psi|^2$, which
is measured at the exit of the interferometer (at $x=10$ from the
barrier) for the two cases (real and imaginary potentials). The
only difference between the two plots is the potential ($i\lambda$
instead of $\lambda$). While the two plots are similar after long
times, their temporal differences are considerably different for
short times as eqs. \ref{short_real} and \ref{short_imag} imply
(like $t^3$ and $t^2$ respectively).

\section{Summary}

The short-time influence of a delta-function potential barrier on
an initially confined wave-packet was investigated. It was shown
that at short times the barrier presence has a negligible
influence, if any, on the wavepacket dynamics. This result applies
also for the probability density of the particles that passed
through the barrier. It was also demonstrated that at short times
an absorptive barrier (i.e. imaginary) has a different impact on
the dynamics than a non-absorptive (i.e., real) one. Namely, at
short times an absorptive barrier appears at the coefficient of
the $t^2$ term, while a non absorptive barrier appears only at the
coefficient of the $t^3$ term. Therefore, it is possible to
distinguish between an imaginary and a real potential barrier by
measuring its effect at short times only on the transmitting
wavefunction. There is no need to measure the transmission and
reflection coefficient simultaneously.

\begin{figure}
\includegraphics[width=8cm,bbllx=30bp,bblly=180bp,bburx=570bp,bbury=620bp]{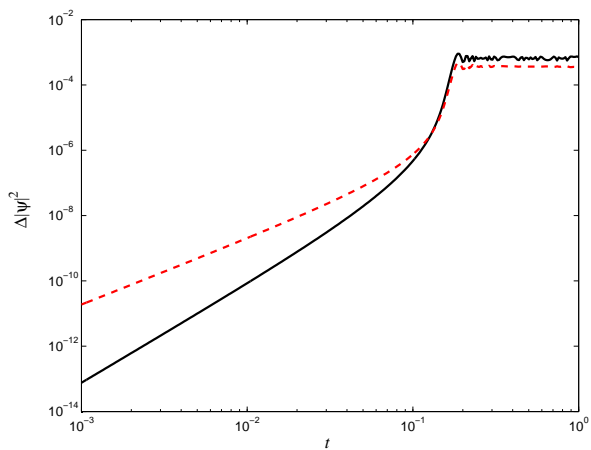}
\caption{\emph{The temporal evolution of $\Delta|\psi|^2$ outside
the interferometer (a distance $x=10$ from the barrier) for a real
potential (solid line) and an imaginary one (dashed line).
$c1=\sqrt{0.49}$ and the other system's parameters are the same as
in fig.2}}\label{fig5}
\end{figure}

\end{document}